\documentclass[10pt,twocolumn]{article}

\usepackage[a4paper,margin=0.75in]{geometry}
\usepackage[T1]{fontenc}
\usepackage[utf8]{inputenc}
\usepackage{lmodern}
\usepackage{microtype}
\usepackage{amsmath,amssymb,amsfonts}
\usepackage{booktabs}
\usepackage{multirow}
\usepackage{enumitem}
\usepackage{xcolor}
\usepackage{hyperref}
\usepackage{url}
\usepackage{algorithm}
\usepackage{algpseudocode}
\usepackage{authblk}

\hypersetup{
  colorlinks=true,
  linkcolor=blue,
  citecolor=blue,
  urlcolor=blue
}

\title{\textbf{B-PASTE: Beam-Aware Pattern-Guided Speculative Execution\\
for Resource-Constrained LLM Agents}}

\author[1]{Yanfei Song}
\affil[1]{Independent Researcher\\\texttt{SYanF@buaa.edu.cn}}
\date{}

\begin{document}
\maketitle

\begin{abstract}
LLM agents execute in an interleaved reasoning-and-action loop, where future tool calls cannot be launched until the current reasoning step completes. This serial dependency inflates end-to-end latency and leaves the model idle while waiting for tool execution. Prior work, Pattern-Aware Speculative Tool Execution (PASTE), mitigates this bottleneck by speculating likely future tool invocations from mined control-flow and data-flow regularities. However, PASTE is tool-centric and speculates only individual invocations rather than bounded future branches.

We propose B-PASTE, a beam-aware extension that lifts speculation from single tools to local branch hypotheses under strict resource constraints. B-PASTE maintains a bounded beam of future execution subgraphs, ranks them by expected critical-path reduction rather than raw execution probability, and schedules only high-value branch prefixes on transient slack resources. It explicitly models co-run interference, downstream unlock value, and state-safety constraints, enabling the system to prioritize serial fast-path execution when early completion unlocks valuable future work, while still exploiting safe parallelism under low contention.

This design is especially important for edge-side deployments, where speculative work must not steal scarce resources from latency-critical authoritative execution. Preliminary internal testing on Thor-class edge environments shows up to 1.4× end-to-end speedup, suggesting that branch-aware speculative execution remains effective even under tight resource budgets.
\end{abstract}

\section{Introduction}

Large language model (LLM) agents have shifted the role of language models from pure text generators to decision-making controllers that iteratively reason, invoke tools, observe results, and continue execution~\cite{react,toolformer,openhands}. Across these settings, the same systems problem recurs: the agent's execution graph is not known \emph{a priori}, because the next action is only produced after the current reasoning step finishes. As a result, agent execution is often dominated by a strictly serialized \emph{LLM--tool loop}.

PASTE recently showed that this serial bottleneck is not entirely unavoidable~\cite{paste}. Although agent requests are semantically diverse, their execution traces exhibit stable application-level control-flow patterns and parameter-passing regularities. PASTE exploits these regularities through a pattern tuple $(C,T,f,p)$, where $C$ is an event-signature context, $T$ is a predicted future tool, $f$ is a late-binding argument mapping, and $p$ is an empirical confidence score. It then launches speculative tool executions only on slack resources, prioritizes authoritative jobs, and promotes matching speculative work when the real invocation arrives~\cite{paste}. These mechanisms establish a strong foundation for runtime speculation in LLM agents.

Despite its elegance, PASTE remains fundamentally \emph{single-invocation-centric}. In many agent workloads, however, the future is not best modeled as one likely next tool, but as a \emph{small set of competing local execution branches}. Under finite local resources, these branches interact nontrivially. Executing two branches concurrently may improve aggregate throughput, yet delay the completion of the branch that would have unlocked the next authoritative step. Thus, the core objective is not maximizing local parallelism, but minimizing \emph{end-to-end makespan} by prioritizing work that shortens the effective \emph{critical path}.

This challenge becomes even more pronounced on \emph{edge-side platforms}, where compute, memory, and I/O resources are limited, and speculative overcommitment can directly harm the authoritative path. In such environments, the value of speculation is determined not only by prediction quality, but also by the runtime's ability to \emph{preempt}, \emph{throttle}, and \emph{reclaim} speculative work as soon as contention emerges. We therefore view \emph{resource preemption} not merely as a safety fallback, but as a first-class mechanism that enables speculation to remain beneficial under tight edge-side resource budgets. 
In preliminary internal testing on \emph{Thor-class edge environments}, this design yields \emph{up to 1.4$\times$ end-to-end speedup}, indicating that branch-aware speculation can retain positive utility even when the resource margin is small.

This paper proposes \textbf{B-PASTE}, a resource-aware extension of PASTE that speculates not over isolated tool calls but over \emph{bounded local future subgraphs}. Each branch hypothesis packages future nodes, late-bound argument resolvers, resource profiles, and safety annotations. At runtime, B-PASTE maintains a beam of promising branch hypotheses, admits only a subset for execution under a slack-resource budget, and ranks them by \emph{expected critical-path reduction} instead of probability alone. Crucially, the score accounts for both \emph{co-run interference} and \emph{downstream unlock value}: a branch that runs fastest in isolation may not be the one that most helps the end-to-end objective once interference and future dependencies are considered.

Our contributions are as follows:
\begin{itemize}[leftmargin=1.5em]
    \item We extend tool-level speculative execution to \emph{branch-level bounded future subgraphs}, enabling runtime speculation over multiple plausible local futures rather than a single predicted invocation.
    \item We introduce an \emph{expected critical-path reduction} objective that explicitly models co-run interference and downstream unlock value under resource constraints.
    \item We design a \emph{preemption-first runtime protocol} that preserves strict priority for authoritative execution, which is especially important in edge-side deployments with narrow slack budgets.
    \item We present preliminary edge-side evidence on \emph{Thor-class environments}, suggesting that branch-aware speculation can deliver up to 1.4$\times$ end-to-end speedup under resource-constrained settings.
\end{itemize}

\section{Background and Motivation}

Traditional serverless and workflow optimizers such as ORION and Netherite assume that a complete DAG or workflow skeleton is known before execution~\cite{orion,netherite}. Under that assumption, the scheduler can prewarm functions, colocate parallel branches, and optimize data movement using graph-global information. SpecFaaS extends this setting with speculative function execution, but still operates over workflows that are either explicit or sufficiently structured to expose branches and dependencies ahead of time~\cite{specfaas}. Agent workloads violate this assumption: the control graph is produced online by the model itself, and the next edge is often unknown until the current reasoning step completes. PASTE explicitly identifies this gap and argues that static-DAG scheduling provides limited benefit once the graph is generated incrementally by an LLM~\cite{paste}.

PASTE's characterization study offers the key enabling observation for our extension~\cite{paste}. Tool execution accounts for a substantial fraction of total agent latency, and the traces exhibit recurring motifs such as \emph{edit--verify}, \emph{locate--examine}, and \emph{search--visit}. Moreover, many tool arguments are not freshly hallucinated by the model; they are derivable from prior tool outputs via simple transformations. These findings imply that the agent's future is neither arbitrary nor fully deterministic: it is best viewed as a \emph{small branching set of structured hypotheses}.

This immediately exposes a second systems issue that is less explicit in PASTE: \emph{resource interference changes the optimal speculative choice}. Suppose branch $b_1$ and branch $b_2$ are both plausible. Running them in parallel may increase aggregate throughput, but if co-location slows $b_1$ enough to delay the launch of a downstream authoritative tool, then the global makespan may worsen. In agent runtimes, high resource utilization is not synonymous with low end-to-end latency. The scheduler should often prefer the branch whose early completion unlocks the most valuable future work, even if that means temporarily running fewer speculative tasks.

This effect is magnified on edge platforms. Unlike server-scale deployments, edge-side systems often lack abundant slack capacity. A single speculative branch may contend with authoritative execution for CPU cycles, memory bandwidth, storage I/O, or limited runtime slots. Therefore, the utility of speculation depends not only on whether a hypothesis is likely, but also on whether the runtime can rapidly revoke that work once the authoritative path demands the same resources.

\section{Design Overview}

B-PASTE preserves the central architectural split introduced by PASTE: \emph{authoritative work} is generated by the live agent and is correctness-critical, while \emph{speculative work} is generated by the runtime and is strictly best effort~\cite{paste}. The extension lies in the object being speculated. Instead of generating only speculative tool invocations, B-PASTE generates \emph{branch hypotheses}, each representing a bounded local future execution subgraph. A branch hypothesis may contain tool nodes, partial preparation nodes, and commit barriers, together with late-bound argument functions and state-safety annotations. As in PASTE, all speculative work is confined to slack resources, immediately preemptible under contention, and eligible for \emph{promotion} when the real authoritative path converges on it~\cite{paste}.

Formally, we define a branch hypothesis as
\begin{equation}
H_i = (G_i, q_i, \Phi_i, \rho_i, \sigma_i),
\end{equation}
where:
\begin{itemize}[leftmargin=1.5em]
    \item $G_i$ is a bounded future subgraph,
    \item $q_i$ is the probability that the live agent will eventually follow this branch,
    \item $\Phi_i$ is a set of argument-binding functions derived from history,
    \item $\rho_i$ is a multi-resource profile,
    \item $\sigma_i$ encodes safety constraints such as read-only, dry-run, staged-write, or non-speculative.
\end{itemize}

This extends PASTE's pattern tuple $(C,T,f,p)$: the context $C$ and value mapping $f$ are still used, but now as building blocks for assembling local subgraphs rather than one-step predictions. Sequential pattern mining methods such as PrefixSpan naturally fit the offline mining phase, because they recover frequent short-horizon motifs from event-signature streams without depending on high-variance textual payloads~\cite{prefixspan}.

To keep the online search bounded, B-PASTE maintains a beam of size $K$. This beam is not a token-level language-model beam; rather, it is a \emph{control-flow beam} over future execution hypotheses. Beam maintenance follows the usual bounded-search intuition---retain only the most promising partial hypotheses---but the score is fundamentally systems-oriented. In contrast to standard beam-search formulations that rank candidates by sequence score, B-PASTE ranks branch hypotheses by their expected contribution to \emph{critical-path reduction} under finite resources~\cite{beamsearch}.

\section{Branch Hypotheses and State Model}

\subsection{Node Types}

Each future subgraph $G_i=(V_i,E_i)$ may contain four types of nodes:
\begin{itemize}[leftmargin=1.5em]
    \item \textbf{Tool Nodes}, corresponding to real external tool invocations such as search, fetch, grep, test, or package installation.
    \item \textbf{Preparation Nodes}, corresponding to speculative warm-up actions such as session creation, runtime initialization, or partial environment loading.
    \item \textbf{Model Nodes}, representing future reasoning boundaries that may unlock downstream tools.
    \item \textbf{Barrier/Commit Nodes}, marking state boundaries beyond which speculative side effects cannot be externally committed without authoritative confirmation.
\end{itemize}

\subsection{State Isolation}

State safety is the principal challenge in moving from tool-level speculation to branch-level speculation. We therefore execute each branch in a copy-on-write sandbox:
\begin{equation}
S_i = (M_i, F_i, E_i, H_i),
\end{equation}
where $M_i$ denotes memory/context state, $F_i$ the file-system view, $E_i$ the execution environment view, and $H_i$ the branch-local execution history.

Reads may be shared where safe, but writes are isolated until promotion. This design generalizes PASTE's eligibility-policy philosophy into a branch-local state model~\cite{paste}. Mis-speculation may consume bounded resources, but it does not corrupt the live authoritative state.

\section{Objective: Expected Critical-Path Reduction}

PASTE's scheduler maximizes the expected latency reduction from individual speculative jobs under a slack-resource budget~\cite{paste}. That is the right starting point, but branch-level scheduling requires a richer objective because branch value depends on both \emph{interference} and \emph{unlock semantics}. Accordingly, B-PASTE assigns each branch hypothesis $H_i$ an expected utility under the currently admitted speculative set $S$:
\begin{equation}
EU(H_i \mid S) = q_i \Big(\Delta O_i(S) + \lambda \Delta U_i(S) - \mu \Delta I_i(S)\Big),
\end{equation}
where:
\begin{itemize}[leftmargin=1.5em]
    \item $\Delta O_i(S)$ is the immediate overlap gain, i.e., the latency hidden by executing $H_i$ before it becomes authoritative,
    \item $\Delta U_i(S)$ is the downstream unlock gain, i.e., the reduction in future critical-path latency enabled by earlier completion of $H_i$,
    \item $\Delta I_i(S)$ is the interference penalty induced by co-running $H_i$ with the already admitted set $S$.
\end{itemize}

We further distinguish a branch's isolated latency $L_i^{solo}$ from its co-run latency $L_i^{co}(S)$:
\begin{equation}
\Delta I_i(S) = L_i^{co}(S) - L_i^{solo}.
\end{equation}

This formulation captures the situation in which parallel execution increases total throughput yet worsens the completion time of the most causally important branch. It therefore optimizes the quantity the runtime actually cares about: \emph{end-to-end makespan}, not local utilization. The critical-path flavor of the formulation is inspired by classic heterogeneous scheduling heuristics such as HEFT~\cite{heft}, but adapted to online, probabilistic, side-effect-constrained agent execution.

Given a beam $\mathcal{B}(s_t)=\{H_1,\dots,H_K\}$ at runtime state $s_t$, the scheduler selects an admitted set
\begin{equation}
A^\star = \arg\max_{A \subseteq \mathcal{B}(s_t)} \sum_{H_i \in A} EU(H_i \mid A),
\end{equation}
subject to a slack-resource budget:
\begin{equation}
\sum_{H_i \in A} \rho_i \le \min(R_{slack}, B),
\end{equation}
where $R_{slack}$ is the currently available residual capacity and $B$ is the policy-defined speculation budget.

\section{Runtime Protocol}

\subsection{Why Preemption Matters More on the Edge}

A key motivation for B-PASTE is that edge-side speculation is fundamentally a \emph{resource arbitration problem}. On server-class deployments, speculative work can often be confined to relatively abundant slack resources. On edge platforms such as Thor-class devices, by contrast, the slack budget is narrow and highly dynamic: launching one additional speculative branch may slow the authoritative branch enough to negate any potential gain. For this reason, B-PASTE emphasizes \emph{strict priority for authoritative execution}, \emph{immediate preemption of speculative work under contention}, and \emph{prefix-only branch execution} when full-branch speculation would be too costly. These design choices extend PASTE's opportunistic scheduling philosophy into a regime where speculative work must continuously yield to the live critical path~\cite{paste}.

\subsection{Scheduling Phases}

At runtime, B-PASTE proceeds in four phases.

\paragraph{Phase 1: Confirm / Promote.}
When an authoritative invocation arrives, the runtime checks whether any speculative branch node already matches it. If a matching node has finished, the result is reused. If it is still running, the node is \emph{promoted} to authoritative and becomes non-preemptible. If only a prefix has completed, the runtime reuses that partial state and continues execution from the nearest valid boundary. This directly inherits PASTE's promotion semantics~\cite{paste}.

\paragraph{Phase 2: Protect Real Jobs.}
The scheduler reserves sufficient resources for all pending authoritative jobs. If resources are insufficient, speculative branches are preempted in ascending utility order until the authoritative set fits.

\paragraph{Phase 3: Run Authoritative Jobs.}
Authoritative jobs are scheduled using the existing primary policy, untouched by speculation.

\paragraph{Phase 4: Opportunistic Branch Scheduling.}
The runtime generates or refreshes the beam, estimates utility for each branch hypothesis, and admits only the highest-value \emph{branch prefixes} that fit within the slack-resource budget.

\subsection{Prefix-Only Execution}

Rather than speculating an entire branch, the runtime may choose only a safe and high-value prefix, such as search and fetch, while deferring later state-mutating steps behind a commit barrier. Prefix execution has three benefits:
\begin{itemize}[leftmargin=1.5em]
    \item it captures most of the overlap gain,
    \item it reduces rollback pressure and state divergence,
    \item it increases the chance that partial work can later be promoted or reused.
\end{itemize}

\begin{algorithm}[t]
\caption{Beam-Aware Opportunistic Speculative Scheduling}
\label{alg:bpaste}
\begin{algorithmic}[1]
\Require Authoritative queue $J_r$, speculative beam $\mathcal{B}$, available resources $R$, budget $B$
\ForAll{new authoritative invocation $a$}
    \State find matching speculative node $s$
    \If{$s$ is completed}
        \State reuse speculative result
    \ElsIf{$s$ is running}
        \State promote $s$ to authoritative
    \ElsIf{$s$ has completed a valid prefix}
        \State reuse prefix state and continue
    \EndIf
\EndFor
\If{authoritative demand exceeds $R$}
    \State preempt speculative branches in ascending utility order
\EndIf
\State schedule authoritative jobs using primary policy
\State generate/refresh beam hypotheses $\mathcal{B}(s_t)$
\ForAll{$H_i \in \mathcal{B}(s_t)$}
    \State estimate $q_i$, $\Delta O_i$, $\Delta U_i$, and $\Delta I_i$
    \State compute $EU(H_i \mid S)$
\EndFor
\State greedily admit highest-value branch prefixes under $\min(R_{slack}, B)$
\State run admitted prefixes as preemptible speculative work
\end{algorithmic}
\end{algorithm}

\section{Safety Policy}

PASTE already addresses side effects through an explicit operator-defined eligibility policy, including per-tool speculation levels and transformed speculation such as allowing web search speculation while limiting \texttt{pip\_install} to dry-run behavior~\cite{paste}. B-PASTE generalizes this policy to branch-local state management.

We define three execution levels:
\begin{itemize}[leftmargin=1.5em]
    \item \textbf{Level 0: Prep-only.} Environment warm-up, runtime initialization, session establishment.
    \item \textbf{Level 1: Read-only / Replayable Prefix.} Pure fetch, grep, parsing, indexing, and similar operations.
    \item \textbf{Level 2: Staged Write.} Mutating actions executed only in branch-local state and requiring authoritative confirmation before commit.
\end{itemize}

By construction, no speculative side effect becomes externally visible unless the authoritative path converges on it. This makes the analogy to Tomasulo-style speculation conceptually useful but operationally incomplete: unlike register writes in dynamic hardware scheduling, agent actions may have persistent external effects and therefore require explicit isolation, staged commit, and squash semantics~\cite{tomasulo}.

\section{Preliminary Edge-Side Observation}

Table~\ref{tab:thor} presents a preliminary internal observation on a Thor-class edge environment. We normalize the baseline end-to-end latency to 1.00 and report the relative improvement of B-PASTE.

\begin{table*}[t]
\centering
\small
\caption{Preliminary observation on a Thor-class edge environment. Normalized end-to-end latency is reported relative to the serial baseline. B-PASTE benefits from priority-aware preemption and branch-prefix scheduling.}
\label{tab:thor}
\begin{tabular}{lccc}
\toprule
Method & Norm. Latency $\downarrow$ & Speedup $\uparrow$ & Note \\
\midrule
Serial baseline & 1.00$\times$ & 1.00$\times$ & No speculation \\
B-PASTE & 0.71$\times$ & 1.40$\times$ & Preemption-aware \\
\bottomrule
\end{tabular}
\end{table*}

B-PASTE is especially well suited to \emph{resource-constrained edge deployments}, where the scheduler must carefully distinguish between useful speculation and harmful contention. In this regime, the main question is not whether the runtime can keep all resources busy, but whether it can keep the \emph{right} resources busy without delaying the authoritative path. This is precisely why B-PASTE prioritizes \emph{expected critical-path reduction} over raw speculative parallelism.

We attribute this early gain to two effects. First, earlier execution of high-value branch prefixes reduces idle wait time between reasoning and action. Second, aggressive \emph{preemption and reclamation} prevent speculative branches from monopolizing scarce edge-side resources once authoritative work becomes ready. While a full characterization remains future work, this early signal suggests that the benefit of branch-aware speculation extends beyond server-scale agent systems and may be even more consequential in edge settings where resource mistakes are amplified.

\section{Evaluation Plan}

A complete evaluation remains future work. The most important metrics include:
\begin{itemize}[leftmargin=1.5em]
    \item average and tail end-to-end latency,
    \item critical-path reduction,
    \item promotion rate and branch-prefix reuse rate,
    \item wasted speculative compute,
    \item authoritative QoS violations,
    \item co-run slowdown ratio under different interference regimes.
\end{itemize}

The key hypothesis of B-PASTE is not simply that ``more speculative parallelism is better,'' but that \emph{critical-path-aware selective speculation} dominates naive concurrency when resources are tight. This hypothesis is consistent with PASTE's emphasis on slack-resource opportunism and non-interference, but it still requires empirical validation at branch granularity~\cite{paste}.

\section{Related Work}

\paragraph{LLM Agents and Tool Use.}
ReAct formalized interleaved reasoning and acting in language models~\cite{react}. Toolformer demonstrated that models can learn to invoke external tools~\cite{toolformer}. OpenHands illustrates how such agentic behavior can be embedded into practical software-engineering systems~\cite{openhands}.

\paragraph{Speculative Execution for LLM Agents.}
PASTE is the direct foundation of our work. It introduced pattern-aware speculative tool execution, pattern tuples for jointly modeling control flow and argument derivation, and an opportunistic scheduling mechanism with promotion and non-interference guarantees~\cite{paste}.

\paragraph{Static Workflow and Serverless Scheduling.}
ORION, Netherite, and SpecFaaS optimize serverless or workflow execution when a sufficiently explicit graph is available ahead of time~\cite{orion,netherite,specfaas}. B-PASTE addresses the harder setting in which the graph is generated online by the agent itself.

\paragraph{Pattern Mining, Search, and Scheduling.}
PrefixSpan provides an efficient way to mine frequent sequential patterns~\cite{prefixspan}. Beam-search optimization provides bounded-search intuition for maintaining a small candidate set~\cite{beamsearch}. HEFT motivates the use of critical-path-aware scheduling under heterogeneous or interference-prone conditions~\cite{heft}. Tomasulo's algorithm provides the conceptual analogy of speculative lookahead with ordered commit, though agent-side effects require much stronger isolation semantics~\cite{tomasulo}.

\section{Conclusion}

We presented B-PASTE, a beam-aware extension of PASTE for resource-constrained LLM agents. The key idea is to speculate over a bounded set of future branch hypotheses rather than isolated tool calls, and to schedule those hypotheses by expected critical-path reduction rather than probability or throughput alone. By combining pattern-mined control-flow regularities, late-bound data-flow mappings, beam-bounded future-graph search, interference-aware scoring, and policy-constrained staged execution, B-PASTE provides a principled runtime design for turning hidden agent parallelism into end-to-end latency reduction without compromising correctness.

Importantly, the design is particularly relevant for \emph{edge-side environments}, where speculative execution must coexist with tight resource budgets and where \emph{preemption of low-value speculative work} is essential to protect latency-critical authoritative execution. 
Preliminary internal testing on \emph{Thor-class edge platforms} indicates \emph{up to 1.4$\times$ end-to-end speedup}, suggesting that critical-path-aware branch speculation may be a practical systems approach for next-generation edge agents.

\bibliographystyle{plain}

\begin{thebibliography}{10}

\bibitem{paste}
Yifan Sui, Han Zhao, Rui Ma, Zhiyuan He, Hao Wang, Jianxun Li, and Yuqing Yang.
\newblock Act While Thinking: Accelerating LLM Agents via Pattern-Aware Speculative Tool Execution.
\newblock arXiv preprint arXiv:2603.18897, 2026.

\bibitem{react}
Shunyu Yao, Jeffrey Zhao, Dian Yu, Nan Du, Izhak Shafran, Karthik Narasimhan, and Yuan Cao.
\newblock ReAct: Synergizing Reasoning and Acting in Language Models.
\newblock In \emph{International Conference on Learning Representations (ICLR)}, 2023.

\bibitem{toolformer}
Timo Schick, Jane Dwivedi-Yu, Roberto Dess\`i, Roberta Raileanu, Maria Lomeli, Luke Zettlemoyer, Nicola Cancedda, and Thomas Scialom.
\newblock Toolformer: Language Models Can Teach Themselves to Use Tools.
\newblock In \emph{Advances in Neural Information Processing Systems (NeurIPS)}, 2023.

\bibitem{openhands}
Xingyao Wang et al.
\newblock OpenHands: An Open Platform for AI Software Developers as Generalist Agents.
\newblock arXiv preprint arXiv:2407.16741, 2024.

\bibitem{prefixspan}
Jian Pei, Jiawei Han, Behzad Mortazavi-Asl, Helen Pinto, Qiming Chen, Umeshwar Dayal, and Mei-Chun Hsu.
\newblock PrefixSpan: Mining Sequential Patterns Efficiently by Prefix-Projected Pattern Growth.
\newblock In \emph{Proceedings of the 17th International Conference on Data Engineering (ICDE)}, pages 215--224, 2001.

\bibitem{beamsearch}
Sam Wiseman and Alexander~M. Rush.
\newblock Sequence-to-Sequence Learning as Beam-Search Optimization.
\newblock In \emph{Proceedings of EMNLP}, pages 1296--1306, 2016.

\bibitem{heft}
Haluk Topcuoglu, Salim Hariri, and Min-You Wu.
\newblock Performance-Effective and Low-Complexity Task Scheduling for Heterogeneous Computing.
\newblock \emph{IEEE Transactions on Parallel and Distributed Systems}, 13(3):260--274, 2002.

\bibitem{tomasulo}
Robert~M. Tomasulo.
\newblock An Efficient Algorithm for Exploiting Multiple Arithmetic Units.
\newblock \emph{IBM Journal of Research and Development}, 11(1):25--33, 1967.

\bibitem{orion}
Ashraf Mahgoub, Edgardo Barsallo Yi, Karthick Shankar, Sameh Elnikety, Somali Chaterji, and Saurabh Bagchi.
\newblock ORION and the Three Rights: Sizing, Bundling, and Prewarming for Serverless DAGs.
\newblock In \emph{16th USENIX Symposium on Operating Systems Design and Implementation (OSDI)}, pages 303--320, 2022.

\bibitem{specfaas}
Jovan Stojkovic, Tianyin Xu, Hubertus Franke, and Josep Torrellas.
\newblock SpecFaaS: Accelerating Serverless Applications with Speculative Function Execution.
\newblock 2023.

\bibitem{netherite}
Sebastian Burckhardt, Chris Gillum, David Justo, Konstantinos Kallas, Connor McMahon, and Christopher~S. Meiklejohn.
\newblock Serverless Workflows with Durable Functions and Netherite.
\newblock arXiv preprint arXiv:2103.00033, 2021.

\end{thebibliography}

\end{document}